\documentclass[]{spie}  

\usepackage{amsmath,amsfonts,amssymb}
\usepackage{graphicx}
\usepackage{rotating}
\usepackage[colorlinks=true, allcolors=red]{hyperref}

\title{Automatic Seizure Prediction using CNN and LSTM}

\author{Abhijeet Bhattacharya}

\affil{Bharati Vidyapeeth's College of Engineering, New Delhi, India}

\begin{document} 
\maketitle

\begin{abstract}
The electroencephalogram (EEG) is one of the most precious technologies to understand the happenings inside our brain and further understand our body's happenings. Automatic prediction of oncoming seizures using the EEG signals helps the doctors and clinical experts and reduces their workload. This paper proposes an end-to-end deep learning algorithm to fully automate seizure prediction's laborious task without any heavy pre-processing on the EEG data or feature engineering. The proposed deep learning network is a blend of signal processing and deep learning pipeline, which automates the seizure prediction framework using the EEG signals. This proposed model was evaluated on an open EEG dataset, CHB-MIT. The network achieved an average sensitivity of 97.746\text{\%} and a false positive rate (FPR) of 0.2373 per hour.
\end{abstract}


\section{INTRODUCTION}
There has been a significant advancement in the field of machine learning through the decades from applications of deep learning from space technologies \cite{9898020, 9378689} to healthcare technologies \cite{bhattacharya}. Many complexes, extensive data, and predictive data analysis can now be approached with a different and unique perspective regarding these new developments and techniques. One such example of such advancements can be noticed in the healthcare sector \cite{freestone2015seizure}.

Epilepsy is a neurological disorder that affects millions of people worldwide \cite{rogers2010r}. People who have epilepsy go through seizures that occur abruptly. The seizures can result in abnormal behavior, anxiety, depression, and even loss of consciousness \cite{tellez2010surgical}. Thus, the unpredictability of the onset of seizures affects the quality of life of these patients adversely. Though one cannot prevent a seizure, predicting the onset of a seizure before it occurs can be a crucial tool for improving epileptic patients' condition. So, to study and diagnose this epileptic seizure, doctors and medical experts monitor the electrical activity going in the brain of the patients who have epilepsy using electroencephalogram (EEG) signals. These EEG signals are detected using small metal
discs with thin wires (electrodes) on the scalp and processing these signals on a computer and recording them. However, this is a complex visual task and requires the doctors' full concentration on these signals to detect any unusual activity in it. Therefore, the prediction of seizures using an algorithm with high sensitivity and low false-positive rate (FPR) can significantly help the patients and doctors take the appropriate steps to reduce the impact of an oncoming seizure. 

Since the past few decades, many researchers have developed different machine learning models and techniques to investigate seizure prediction using EEG data \cite{gadhoumi2016seizure, kuhlmann2018seizure}. Early approaches included working on a particular feature or a combination of features to detect the seizure before its onset. The issue with this approach was the inconsistency of the model as a whole. Though the model worked on some particular cases, it could not adjust itself dynamically to determine an oncoming seizure. Thus, it could not be accepted as a general bias for seizure prediction. Another major issue while solving the problem of epileptic seizure prediction is the unavailability of the data sets. A working prediction algorithm with sufficiently high sensitivity and great efficiency requires long hours of EEG signal data. Thus, various models and techniques were proposed for an efficient and reliable seizure prediction algorithm; the problem remains unsolved.

\begin{figure}
  \centering%
    \graphicspath{ {./} }
    \includegraphics[width=8cm,height=8cm]{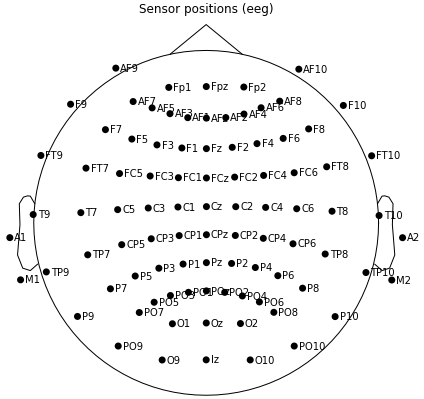}
    \caption{\label{fig:sensors_pos}All the sensors used to capture EEG signals according to the standardized placements}
\end{figure}

With the advancements in big data analysis and deep learning, newer approaches came, which favored this problem of seizure prediction and made it easy to solve \cite{bhattacharya}. The hundreds of hours of complex EEG signals could be analyzed with higher efficiency. Moreover, feature extraction of higher levels could be analyzed, and hidden patterns could be detected between the signals. All of these factors favored Convolutional Neural Networks (CNN) to predict epileptic seizures \cite{krizhevsky2017imagenet}. This is because CNN's are extensively used in computer vision, image processing, and EEG classification. It can be used to observe the spike in the EEG signals, which are a breakthrough for detecting seizures. These algorithms were significantly better than the first approaches and gave much higher sensitivity and low False Positive Rate (FPR). Though the CNN algorithm results were significantly higher, there were still issues with the features' inconsistency. For the time-series data, CNN tends to underperform as compared to other visual tasks. Because of the high signal-to-noise ratio, the CNN gets degraded as general and fails to detect some important feature. Another study used Long Short-Term Memory (LSTM), a deep learning model. This model improves the existing Recurrent Neural Networks (RNN) because of the different gates that control the learning rate \cite{petrosian2000recurrent, graves2012supervised}. These gates are the input gate, the output gate, and the forget gate. The design of the LSTM model helped in determining the temporal characteristics of the EEG signals. This is because the EEG signals are complex and dynamic.

\begin{figure}
  \centering%
    \graphicspath{ {./} }
    \includegraphics[width=15cm,height=5cm]{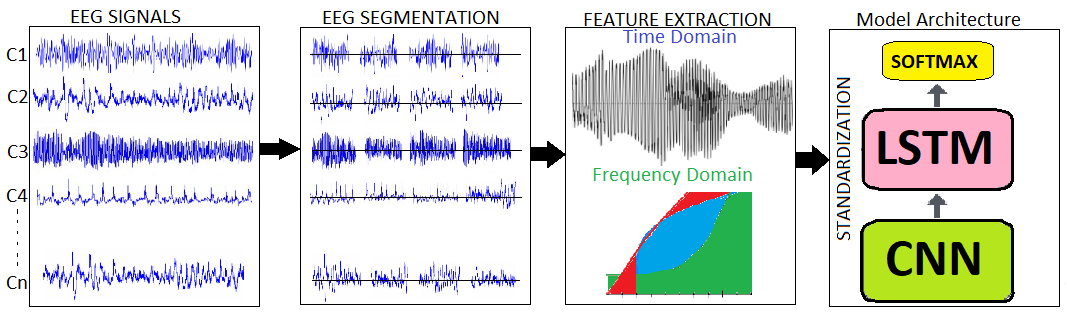}
    \caption{\label{fig:block_diagram}Block diagram of our proposed work}
\end{figure}

Though LSTM models had some advantages over the CNN model, it can be observed that the higher-order data analysis provided by the CNN models could still be put to good use. The defined seizure prediction horizon (SPH) in the earlier model using LSTM was set to zero \cite{maiwald2004comparison}. This is not ideal in a seizure prediction system for the patients, and hence, the seizure occurrence period (SOP) will have to be manually adjusted, which is not good for a reliable seizure prediction algorithm. In this study, a temporal CNN + LSTM model was introduced for seizure detection in EEG signals. The proposed model enriched the features extracted by the temporal CNN, which boosted the features and made them dynamic. The end-to-end proposed pipeline is an apt model for the time-series data with its features in the time domain and frequency domain. Refer to Figure \ref{fig:block_diagram} for a block diagram of our proposed work.

Discussion of different techniques used, features extracted, preictal duration, setting value for SPH, and many more is done in the coming sections. Section 2.1 showcases detailed information about the datasets; section 2.2 describes the general properties of an EEG signal, section 2.3 describes the different voltages in an EEG signal, section 2.4 describes the method used in our work to extract the features from the EEG signals, section 2.5 describes the architecture used in this work for our end-to-end pipeline, section 2.6 describes the training schemes and methods used in this work, section 2.7 describes the system evaluation of our proposed method, section 2.8 describes the post-processing method used, section 3 describes the results of our model, section 4 discusses the seizure prediction and other methods used before, section 5 concludes our work. 

\section{Methods}
\subsection{CHB-MIT Scalp EEG Database}
All the EEG signals and data used in this work were taken from the open Boston Children's Hospital (CHB)-MIT dataset \cite{goldberger2000physiobank}, which is publically available at PhysioNet.org \cite{chbmit}. The dataset contains the EEG signals collected from 24 pediatric subjects where there were seventeen females, ages from 1.5 to 19 years; five males, ages from 3 to 22 years; one missing gender/age data with intractable seizures at the Children's Hospital Boston. All the 24 case subjects in this dataset were recorded at a sampling frequency of 256 samples per second or a sampling rate of 256 Hz, and all these EEG recordings were done using the International 18-30 electrodes position system with a 16-bit resolution from them refer to Figure \ref{fig:sensors_pos}. This dataset contains 198 seizures recorded on continuous EEG recordings with 983 hours of epileptic seizures. In almost all EEG recorded data in the dataset, the EEG signals are segmented in 1-hour long epochs, but some files are 2-4 hours long. After visual inspection, the clinical experts at the Children's Hospital recorded a total of 198 seizures at the Children's Hospital, for which they manually annotated the seizure onset and offsets, time intervals for any EEG epochs with ictal activity. This dataset has many seizures confined in small channels and .edf files. Due to such a high density of data in these channels and files, multiple seizures occur close to each other. Refer to Table \ref{table:chbmit} for further details of the CHBMIT dataset. \\

In predicting the seizures, the leading seizure was the target to detect and raise the alarm. Accordingly, for any two back-to-back seizures that happen with a delay of under 30 minutes, those are considered one single seizure and utilized at the start of the principal seizure as the beginning of the joined seizure. According to the annotated files in the dataset, in mostly all the subjects, the EEG recording montages keeps changing with time; some channels were constant throughout every patient while some channels were being continuously added or removed during the recording process. Due to this constant change in every patient's channels in the dataset, only those channels are selected for analysis, which is constantly available throughout the duration to mitigate data heterogeneity. Therefore, 18 channels are common in all the subject's data, namely: C3-P3, C4-P4, CZ-PZ, F3-C3, F4-C4, F7-T7, F8-T8, FP1-F3, FP1-F7, FP2-F4, FP2-F8, FZ-CZ P3-O1, P4-O2, P7-O1, P8-O2, T7-P7, and T8-P8. From all the 24 patients, certain patients were discarded from the dataset, namely: Patient 12, because it did not have certain channels from the names of the above channels, Patient 07 because it had very inconsistent data. Thus, removing all this data from the available dataset, the final dataset consists of about 867 hours of continuous EEG signals of data with 147 seizures from 21 cases.

\begin{figure}
  \centering%
    \graphicspath{ {./} }
    \includegraphics[width=12cm,height=5.5cm]{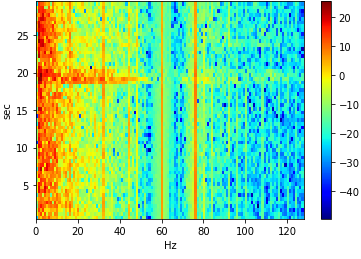}
    \includegraphics[width=12cm,height=5.5cm]{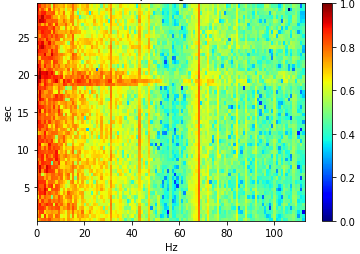}
    \includegraphics[width=12cm,height=7cm]{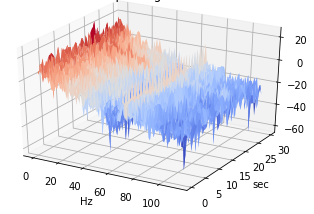}\\
    \caption{\label{fig:spectrograms}From top to bottom (a) Example short-time Fourier transform spectrogram of a 30-s window. (b) Same spectrogram after removal of line noise and standardization. (c) A 3-D representation of cleaned spectrogram }
    
\end{figure}

\subsection{EEG features}
The electrical activity recording in the encephalon from the scalp is called the electroencephalogram (EEG). The recorded EEG signals are the reflection of the cortical electrical activity. Signal intensity: EEG activity is quite minuscule, quantified in microvolts (mV). Signal frequency: The EEG waves of human with main frequencies are:
\begin{itemize}
\item Delta(\(\delta\)): It has a frequency of 3 Hz or below. This kind of EEG signal, by and large, relates to the most elevated in amplitude and the slowest waves. For one year or in stages 3 and 4 of rest, this element is ordinary as infants beat ascendant beat. This is expectedly perceptible at the back in kids and frontally in grown-ups.

\item Theta(\(\theta \)): The frequency of this EEG includes lies between 3.5 to 7.5 Hz and is consigned as "slow" action. It is normal in kids up to 13 years and in sleep yet whimsical in stimulated grown-ups. It may be outwardly seen as a sign of major subcortical injuries; it can be optically observed in the summed up dispersion in diffuse issues, such as metabolic encephalopathy or a few occasions hydrocephalus.

\item Alpha(\(\alpha \)): This feature's frequency lies in 7.5 Hz to 13 Hz. It has a higher sufficiency on the head's ascendant side yet is best optically recognized on each side of the back locale in the head. It shows up when shutting the visual perceivers and unwinding and disappears when opening the visual perceivers or cautioning by any instrument (intellectually imagining, figuring). Commonplace grown-ups who are loose have this element optical perceived as a significant beat. It is available the entire life; however, it gets unmistakable after the thirteenth year.

\item Beta(\(\beta \)): beta activity is "expeditious" activity. It has a frequency of 14 and more prevalent Hz. It is generally optically observed on the two sides in balanced dispersion and is the most prominent frontally. It very well might be missing or shortened in regions of cortical harm. It is by, and largely viewed as an unremarkable cadence. It is the ascendant musicality in patients who are careful or worried or have their visual perceivers open.

\item Gamma(\(\gamma \)): Gamma Activity (GBA) Gamma-band action involves an EEG frequency range, from 30 to 200 Hz, and is appropriated generally all through cerebral structures. GBA participates in various cerebral capacities, for example, perception, attention, recollection, consciousness, synaptic plasticity, and motor control.

\end{itemize}
  
It tends to be seen that there is an extraordinary promising possibility in distinguishing preictal varieties in the scalp and intracranial signs by the supreme contrasts in power in every one of these frequency bands \cite{park2011seizure, zhang2015low, bandarabadi2015epileptic, truccolo2011single}. Frequency refers to a perpetual rhythmic activity (in Hz). There are various properties shown by the recurrence of EEG movement; these are: 
 \begin{itemize}
  \item Rhythmic. Influxes of consistent recurrence are there in EEG action.
  \item Arrhythmic. There is no steady rhythm available in the EEG movement.
  \item Dysrhythmic. Rhythms and/or patterns of EEG activity that characteristically appear in patient groups or infrequently or visually perceived in salubrious subjects.
\end{itemize}

\begin{table}[ht]
\caption{\label{table:chbmit}Details of the CHB-MIT Scalp dataset. }
\begin{center}
\begin{tabular}{ |p{2.3cm}|p{1cm}|p{1cm}|p{2cm}|p{2cm}| p{2cm}|}
 \hline
 \textbf{ Patient number} & \textbf{Sex} & \textbf{Age (years)} & \textbf{Number of EEG channels} & \textbf{Number of seizures} & \textbf{Length of Recordings (hh:mm:ss)}\\
 \hline
 1 & F & 11 & 18 & 7 & 40:33:08\\
 2 & M & 11 & 18 & 3 & 35:15:59\\
 3 & F & 14 & 18 & 7 & 38:00:06\\
 4 & M & 22 & 18 & 4 & 156:03:54\\
 5 & F & 7 & 18 & 5 & 39:00:10\\
 6 & F & 1.5 & 18 & 10 & 66:44:06\\
 7 & F & 14.5 & 18 & 3 & 67:03:08\\
 8 & M & 3.5 & 18 & 5 & 20:00:23\\
 9 & F & 10 & 18 & 4 & 67:52:18\\
 10 & M & 3 & 18 & 7 & 50:01:24\\
11 & F & 12 & 18 & 3 & 34:47:37\\
13 & F & 3 & 18 & 12 & 33:00:00\\
14 & F & 9 & 18 & 8 & 26:00:00\\
15 & M & 16 & 18 & 20 & 40:00:36\\
16 & F & 7 & 18 & 10 & 19:00:00\\
17 & F & 12 & 18 & 3 & 21:00:24\\
18 & F & 18 & 18 & 6 & 35:38:05\\
19 & F & 19 & 18 & 3 & 29:55:46\\
20 & F & 6 & 18 & 8 & 27:36:06\\
21 & F & 13 & 18 & 4 & 32:49:49\\
22 & F & 9 & 18 & 3 & 31:00:11\\
23 & F & 6 & 18 & 7 & 26:33:30\\
24 & – & – & 18 & 16 &  21:17:47\\
 \textbf{\textit{Total:}}& & & & \textbf{\textit{158}} & \textbf{\textit{979:56:07}}\\
 \hline
\end{tabular}\\
\end{center}
\end{table}

\subsection{EEG Voltages}
EEG voltage alludes to the pinnacle voltage of the EEG movement. Qualities are reliant, partially, on the accounting procedure. Spellbinding terms related to EEG voltage include: 

\begin{itemize}
  \item \textbf{Attenuation} (equivalents: concealment, misery). Decrease of the sufficiency of EEG movement occurring because of reduced voltage. Precisely when the gesture diminishes movement, it is said to have been "blocked" or to show "blocking."
  
  \item \textbf{Hypersynchrony}. Seen as an expansion in voltage and conventionality of cadenced turn of events, or inside the alpha, beta, or theta go. The term proposes advancement in the number of neural parts adding to the beat. (Note: the term is utilized in an interpretative sense, in any case, as a descriptor of progress in the EEG).
  
  \item \textbf{Paroxysmal}. The development emerges from the establishment with a quick start, coming to (conventionally) exceptionally high voltage and fulfillment with a surprising re-visitation of cutting down voltage activity. Even though the term does not construe anomaly, much uncommon development is paroxysmal.
\end{itemize}

\subsection{Feature extraction}
The feature extraction algorithm must extract the important features in a loss-less manner, theoretically, but the extraction must practically retain the most important features from the EEG epochs. Most prevalent features were extracted from each of the 30 seconds segments of EEG signals to feed the data into the model as input. The EEG signal's raw format was very computationally expensive due to its continuous time-varying form, so discretizing these EEG signals was done \cite{gadhoumi2012discriminating, wang2011temporal}. The most efficient way to process this raw EEG data is to convert them into the time domain and frequency domain \cite{direito2011feature}. Hence, the easiest and convenient way to convert this time-varying data into the frequency domain is to use the Fourier-Transform \cite{rasekhi2013preprocessing, greaves2014predicting}, and the Wavelet-transforms \cite{direito2011optimized, moghim2011evaluating} more categorically Expeditious Fourier Transform, FFT, and Discrete Wavelet Transform, DWT. This study included the short-time Fourier transform (STFT) as shown in equation 1, to get the frequency domain from the time-varying EEG data, and then this frequency domain data was clubbed with the time domain data to form the Spectrogram dataset from the 30 seconds continuous data. However, these EEG raw recordings were contaminated with power line noises. This problem was solved by removing the frequency components ranging from 57–63 Hz and 117–123 Hz in the frequency domain. After cleaning and normalizing this Spectrogram data, this Spectrogram was saved in the NumPy format in the matrix shape of 18*59*114, 18 representing the number of EEG signal channels patient's data, 59*114 represents the data in the frequency domain, and time domain. For further visualization of the spectrograms and the STFT of a 30-s window after abstraction of puissance line noise, visually perceive Figure \ref{fig:spectrograms}.
 
 \begin{equation}
\textbf{STFT }y(t) (w,{\displaystyle \tau })  = \int_{-\infty}^{\infty} y(t){\displaystyle w}(t - {\displaystyle \tau })e\mathbb{}^{-iwt} dt
\end{equation}
Here y(t) is the signal to be transformed, $w(\tau)$ \text{is Gaussian window function}\\
 
The most common problem in machine learning of highly imbalanced datasets, i.e., more training data in one class than in others \cite{wang2011temporal}, was also present in the dataset. Thus more of the preictal segments had to be generated by the overlapping sampling technique during the training phase. Hence these additional examples were made by sliding the 30 seconds window at each progression \textbf{S} throughout preictal time-series EEG signals. Here \textbf{S} is a variable and is carefully chosen according to the patient-patient so that there remain the same number of samples per class in the training dataset.

\subsection{Architecture}
\begin{figure}
  \centering%
    \graphicspath{ {./} }
    \includegraphics[width=15cm,height=5cm]{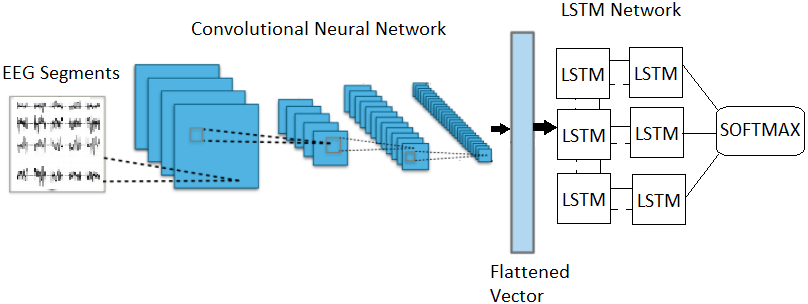}
    \caption{\label{fig:model_arch}Schematic representation of the proposed model architecture}
 \end{figure}

 Convolutional neural network (CNN) and Long short term memory (LSTM) have been extensively used in the field of computer vision
 and natural language processing. They are even used for this problem statement of seizure prediction. In this work, an architecture combining both CNN and LSTM is introduced. The architecture of the proposed method is shown in Figure \ref{fig:model_arch}. Windows of fixed-length are slid through the input EEG signal to extract the series of EEG segments, further passed into a Fully Convolutional Neural Network (FCNN). The output of the FCNN is further given as an input into the Long-short term memory(LSTM) network. Lastly, a fully connected and a softmax layer is utilized to attain each class's probability, which helps classify the oncoming input EEG segment.

 \subsubsection{Fully convolutional network}
 Time-series signals like EEG signals are analyzed extensively by the Temporal convolutional neural network as they can apprehend the internal learning of structures of these EEG signals. Characteristics of low frequency with long periods and high-frequency characteristics with short periods are present in raw EEG signals. So, these features have a hierarchy among them. This hierarchy of features can be learned automatically with a fully convolutional network and capturing the data's internal structure. Furthermore, a low-level layer learns miniature features, while a higher-level layer learns high-level features.
\begin{equation}
    A[b,c] = (d*e)[b,c] = \sum_{x}\sum_{y}e[x,y]d[b-x,c-y]
\end{equation}
d denotes the input image, and e denotes the kernel; b and c denote the indexes of rows and columns of the resultant matrix.

Feature extraction is done using a temporal convolutional network, and it has been substantiated to be an effective method for problems related to time-series. A simple yet effective FCNN architecture is considered for training the model, which helps avoid overfitting. It includes three identically connected convolutional blocks. Each of the three basic blocks comprises a convolutional layer with a Rectified Linear Unit (RELU) as its activation function. The number of kernels for the three convolution layers is 20, 40, and 60, respectively.  The first convolution layer has a kernel size of 7*7 and a stride of 2*2. The second and third layers have kernel size 7*7 for both layers and a stride of 1*1 and 2*2. The output of the FCNN model is given to the Long short-term memory unit.

\subsubsection{Long short-term memory}
Neurologists determine if an EEG epoch belongs to the seizure part by checking how close or far is the EEG recording from the current epoch. Recurrent neural networks (RNN) has made compelling advancement to be able to replicate results like human beings. So, RNN based architectures can identify the current EEG signal by taking advantage of the knowledge learnt previously. However, RNNs tend to perform poor because of the vanishing gradient problem. This led to the formation of an advanced architecture called Long short-term memory (LSTM). It includes memory mechanism, which helps in mitigating the vanishing gradient problem. This memory mechanism assists the model to preserve the previous information of the EEG recordings. The equation of the LSTM are as follows:
\begin{equation}
    \tilde{i}_t= \tilde{\sigma}_i(\tilde{x}_t \tilde{W}_{xi} \tilde{h}_{t-1}) \tilde{W}_{hi} + \tilde{b}_i),
\end{equation}
\begin{equation}
    \tilde{f}_t= \tilde{\sigma}_f(\tilde{x}_t \tilde{W}_{xf} \tilde{h}_{t-1}) \tilde{W}_{hf} + \tilde{b}_f),
\end{equation}
\begin{equation}
    \tilde{c}_t = \tilde{f}_t \odot \tilde{c}_{t-1} + \tilde{i}_t \odot \tilde{\sigma}_c(\tilde{x}_t \tilde{W}_{xc} +\tilde{h}_{t-1} \tilde{W}_{hc} + \tilde{b}_c),
\end{equation}
\begin{equation}
    \tilde{o}_t= \tilde{\sigma}_o(\tilde{x}_t \tilde{W}_{xo} \tilde{h}_{t-1}) \tilde{W}_{ho} + \tilde{b}_o),
\end{equation}
\begin{equation}
    \tilde{h}_t = \tilde{o}_t \odot \tilde{\sigma}_h(\tilde{c}_t),
\end{equation}
where $\tilde{x}_t$ and $\tilde{h}_{t-1}$ are the inputs to the inner LSTM unit and can be computed with the aid of outer unit parameters:
\begin{equation}
    \tilde{x}_t= i_t \odot{\sigma}_c(x_t W_{xc}+h_{t-1} W_{hc} + b_c),
\end{equation}
\begin{equation}
    \tilde{h}_{t-1} = f_t \odot \tilde{c}_{t-1},
\end{equation}
where the three gates are illustrated by $\tilde{}{i}_t$, $\tilde{f}_t$ , and $\tilde{o}_t$; input cell state is illustrated by $\tilde{c}_t$; the three gates and cell input is connected to $\tilde{x}_t$ by weight vectors $\tilde{W}_{xi}$ , $\tilde{W}_{xf}$ , $\tilde{W}_{xo}$ , and $\tilde{W}_{xc}$; the three gates and cell input is connected to $\tilde{h}\textsubscript{t-1} $ by weight vectors $\tilde{W}_{hi}$ , $\tilde{W}_{hf}$ , $\tilde{W}_{ho}$ , and biases of the three gates and cell input are illustrated by $\tilde{W}_{hc}$; $\tilde{b}_i$ , $\tilde{b}_f$ , $\tilde{b}_o$ , and $\tilde{b}_c$; sigmoid function is denoted by $\sigma$, and scalar product between two vectors is denoted by $\cdot$. \\\\
The output of the inner LSTM unit varies the call state of the outer LSTM, and it is given as
\begin{equation}
    c_t= \tilde{h}_t
\end{equation}


\subsection{System evaluation}
Before moving onto metrics to test the study, two additional parameters called seizure prediction horizon (SPH) and seizure occurrence periods (SOP) are introduced for the system evaluation. We considered these two factors, according to \cite{maiwald2004comparison}. The time interval where an oncoming seizure is expected to occur is called the SOP factor. The time duration between the alarm of an incoming seizure and the start of an epileptic seizure or SOP is known as SPH. For an incoming seizure, if the alarm for the seizure falls in the SPH period, and the seizure occurs in the SOP period, it is classified as a perfect seizure prediction or true positive seizure. Similarly, if an alarm is raised in the SPH period, but there is no seizure in the SOP period, it is classified as a false positive seizure. 

According to EEG experts, the SOP time interval must not be too long, which causes the patient's anxiety to rise, and the SPH factor must be long enough to sufficient intervention or precautions (SPH is also called intervention time \cite{gadhoumi2016seizure}). In this study, the SOP time and the SPH factor are fixed to be 30 minutes and 3 minutes, respectively. The SOP, FPR metrics are used to predict the incoming seizure according to equation 12 given by \cite{schelter2006testing}:
\begin{equation}
P \approx 1 - e^{-FPR\cdot SOP}\\
\end{equation}
Therefore the probability of predicting at least
k of L independent seizures by chance is given by equation 13:\\
\begin{equation}
P = \displaystyle\sum_{i\geq k} \begin{bmatrix} 
		L & i
	\end{bmatrix} P^{i}(1-P)^{L-i}
\end{equation}

\subsection{Post-Processing}
In general, it is a typical practice to separate bogus positives during interictal periods. In work done by Park et al., 2011 \cite{park2011seizure}, they recommended that these disconnected bogus expectations be viably diminished by utilizing a discrete-time Kalman filter. In this research work, a basic technique is utilized called k of n, in which a crisis caution is set when out of n seizure forecasts, k seizure expectations were positive. For model engineering, k = 8 and n = 10 were chosen, making a decent decision for a proficient forecast. This implies if during the last 300 s in any event, 240 s resulted in a positive prediction, at that point, the caution is set.

\begin{table}[t]
\caption{\label{table:results}Results of our proposed seizure prediction algorithm on the CHB-MIT EEG dataset. SEN = Sensitivity.}
\begin{center} 
\begin{tabular}{ |p{1cm}|p{2cm}|p{2cm}|p{2cm}|p{2cm}| }
 \hline
 \textbf{Case} & \textbf{Number of seizures} & \textbf{EEG duration (hrs)} & \textbf{SEN} (\text{\%}) & \textbf{FPR} ($h^{-1}$)\\
 \hline
 1 & 7 & 40.5 & 100 & 2.80\\
 2 & 3 & 35 & 100 & 0\\
 3 & 7 & 38 & 100 & 0\\
 4 & 4 & 156 & 100 & 0\\
 5 & 5 & 39 & 100 & 0\\
 6 & 10 & 66.5 & 97.560 & 0.6\\
 8 & 5 & 20 & 99.3240 & 0\\
 9 & 4 & 68 & 100 & 0\\
 10 & 7 & 50 & 100 & 0\\
 11 & 3 & 35 & 100 & 0\\
 13 & 12 & 33 & 100 & 0\\
 15 & 20 & 40 & 98.663 & 0.1560224\\
 16 & 10 & 19 & 98.78 & 0\\
 17 & 3 & 21 & 100 & 0.89\\
 18 & 6 & 35.5 & 100 & 0.0571428\\
 19 & 3 & 30 & 100 & 0\\
 20 & 8 & 27.5 & 100 & 0\\
 21 & 4 & 33 & 100 & 0.085714\\
 22 & 3 & 31 & 67 & 0.366\\
 23 & 7 & 26.5 & 93 & 0\\
 24 & 16 & 21 & 98.333 & 0.0280\\
 \hline
 \textbf{\textit{Total:}} & & & \textbf{\textit{97.746}} & \textbf{\textit{0.2373}}\\
 \hline
\end{tabular}\\
\end{center}
\end{table}

\section{Results}
 All the metrics in this study were calculated using an SOP duration of 30 minutes with an SPH factor of 3 minutes. With the help of all the previous theories, the EEG segments of the interictal class were divided into more minute subgroups while maintaining the size identical to that of the preictal class to solve the problem of class imbalance. Leave-one-out cross-validation was performed twice, and all the average results with their standard deviation were noted. All these seizure results are compiled in Table \ref{table:results}. Mainly, two metrics are considered for evaluating the proposed methodology: Sensitivity and False positive rate (FPR). FPR is the rate of false alarms as the number of false positives per hour of EEG recordings. Four sub-metrics were calculated to find sensitivity and FPR. These are :

\begin{enumerate}
\item \textbf{True positive (TP): } The true positive predictions are those predictions by the model that are correctly classified as preictal EEG samples.
\item \textbf{True negative (TN): } The true negative predictions are those predictions by the model that are correctly classified as interictal EEG samples.
\item \textbf{False positive (FP): } The false-positive predictions are those predictions by the model that are incorrectly classified as preictal EEG samples.
\item \textbf{False negative (FN): } The false-negative predictions are those predictions by the model that are incorrectly classified as interictal EEG samples.
\end{enumerate}

So, sensitivity can be calculated by TP/(TP + FN), and FPR can be calculated as the number of FP's in one hour.\\

The evaluation of the proposed model for seizure prediction on the test data displayed extremely positive results. It performed well on all the 158 seizures present in the benchmark dataset. The average FPR achieved was 0.2373 per hour, and the average sensitivity of the proposed model was 97.746\text{\%}, as seen from Table \ref{table:results}. Furthermore, 20 out of the 24 patients attained zero false alarms by the proposed model in the CHB-MIT database. This shows a high caliber of seizure prediction even under varying conditions. Thus, helping us achieve practical clinically-based seizure prediction.

Using the above-stated metrics, the proposed method achieved an average sensitivity and FPR of 97.746\text{\%} and 0.2373 per hour, respectively.


 \begin{figure}
  \centering%
    \graphicspath{ {./} }
    \includegraphics[width=13cm,height=6cm]{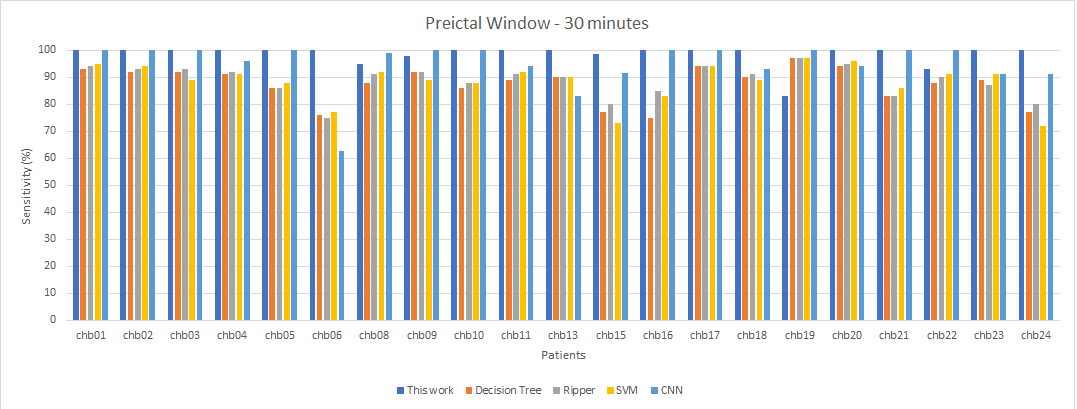}
    \caption{\label{fig:comparision}Comparision of our model's performance with other models on sensitivity}
\end{figure}

\begin{sidewaystable*}
\caption{\label{table:banchmark}Benchmarking of recent seizure prediction approaches and this work. Here FB = Freiburg Dataset, MIT = CHB-MIT dataset, A.E.S. = American Epilepsy Society Seizure Prediction Challenge Dataset}
{\begin{tabular}{@{}cccccccccc@{}}

\\\hline\\[-6pt]
Authors & Dataset & Feature & Classifier & Seizures & Sensitivity (\text{\%}) & FPR (/h) & SOP & SPH\\
\hline\\[-6pt]
 Cho et al. \cite{cho2016eeg} & MIT, 21 patients &  Phase locking value & SVM & 65 & 82.44 & - & 30 min & 3 min\\

 Maiwald et al.\cite{maiwald2004comparison} & FB, 21 patients & Dynamical similarity & Threshold crossing & 88 & 42 & $\leqslant$ 0.15 & 30 min & 2 min\\
 
 Winterhalder et al.\cite{winterhalder2006spatio} & FB, 21 patients & Phase coherence & Threshold crossing & 88 & 60 & 0.15 & 30 min & 10 min\\
 
  Park et al.\cite{park2011seizure} &FB, 18 patients & Univariate spectral power & SVM & 80 & 98.3 & 0.29 & 30 min & 0h\\
 
 Li et al.\cite{li2013seizure} & FB, 21 patients & Spike rate & Threshold crossing & 87 & 72.7 & 0.11 & 50 min & 10 s\\
 
 Zheng et al.\cite{zheng2014epileptic} & FB, 10 patients & Mean phase coherence & Threshold crossing & 50 & $\geqslant$ 70 & $\leqslant$ 0.15 & 30 min & 10 min\\
 
 Eftekhar et al.\cite{eftekhar2014ngram} & FB, 21 patients & Multiresolution N-gram & Threshold crossing & 87 & 90.95 & 0.06 & 20 min & 10 min\\

  Aarabi \text{\&} He et al.\cite{aarabi2017seizure} & FB, 21 patients & Bayesian inversion & Rule-based decision & 87 & 87.07 & 0.20 & 30 min & 10 s\\
&&of power density&&&&&\\
 Zhang \text{\&}Parhi\cite{zhang2015low} & FB, 18 patients & Power spectral density & SVM & 80 & 100 & 0.03 & 50 min & 0h\\

  Zhang \text{\&}Parhi\cite{zhang2015low} & MIT, 17 patients & Power spectral density & SVM & 76 & 98.68 & 0.05 & 50 min & 0h\\

Parvez \text{\&}Paul\cite{parvez2016seizure} & FB, 21 patients & Phase-match error,& LS-SVM & 87 & 95.4 & 0.36 & 30 min & 0h\\
&&deviation, fluctuation&&&&&\\
Aarabi \text{\&} He\cite{aarabi2014seizure} & FB, 10 patients & Univariate and bivariate  & Rule-based decision & 28 & 86.7 & 0.126 & 30 min & 10 s\\

Khan et al.\cite{khan2017focal} & MIT, MSSM & Wavelet transform & CNN & 131 & 87.8 & 0.14 & 10 min & 0h\\

Truong et al.\cite{truong2018convolutional} & MIT, 13 patients & Short-time Fourier transform & CNN & 64 & 81.2 & 0.16 & 30 min & 5 min\\

Sharif \text{\&} Jafari\cite{sharif2017prediction} & FB, 19 patients & Fuzzy rules on Poincar'e plane & SVM & 83 & 91.8–96.6 & 0.05–0.08 & 15 min & 2-42 min\\

Liu et al.\cite{liu2020automatic} & FB, 21 patients & S-transform & CNN & 66 &  97.01 & 0.36 &  & \\

Y. Li et al.\cite{li2018epileptic} & MIT, 21 patients & - & FCNLSTM & 135 &  94.07 & 0.66 &  & \\

This work & MIT, 21 patients & Short-time
Fourier transform & Transformer Model & 158 & 97.746 & 0.2373 & 30 min & 3 min\\
\hline
\end{tabular}\label{tab2}}
\end{sidewaystable*}

\section{Discussion}
The conventional algorithms solved seizure prediction task by handpicking some of the selected features from the EEG continuous signals and constantly studying them. These manually handpicked features were generally less. They were dependent on classical methods for predicting an oncoming seizure, for example, spike rate, mean phase coherence, power spectral density ratio, etc., resulting in poor results for reasons such as handcrafted features, low dimensional data, dependence on low evaluating metrics. Machine learning algorithms came into the picture to handle high dimensional data and advanced evaluating metrics for an oncoming seizure to deal with these problems.
A given EEG segment was consigned as interictal or preictal by Cho et al. \cite{cho2016eeg} with the help of an SVM classifier. This was done by calculating phase-locking values between scalp EEG signals by cumulating multivariate empirical mode decomposition. The CHB-MIT database comprised of 65 seizures, and their methodology was employed on all these 65 seizures. They secured a sensitivity of 82.44\text{\%} and specificity of 82.76\text{\%}. The initial set of features were extracted by performing spectral analysis on scalp EEG signals by Zhang and Parhi \cite{zhang2015low}. This benefited their SVM classifier to get the most suitable features and electrodes, which helped them differentiate between preictal and interictal EEG segments. This model was evaluated on an EEG dataset covering 78 seizures. A sensitivity of 98.68\text{\%} and a False Positive Rate of 0.05 per hour was attained by this method. Other machine learning algorithms like the k-Most nearest Neighbours classifier was additionally evaluated on multi-day recordings of 10 patients but resulted in remotely lower performance with a sensitivity of 73\text{\%} and specificity of 67\text{\%}.
\\
If the comparison between the proposed model or, in general, deep learning algorithms with the conventional algorithms of machine learning like rule inducers, Decision Tree, and SVM is to be done, then it can be inferred that the deep learning algorithms have surely affected more in seizure prediction. 
With the addition of neural networks and deep learning algorithms to this problem, the model learns the EEG signals' features, thereby making the whole architecture more efficient and accurate. Especially, CNN has grabbed the maximum attention in epilepsy detection and prediction of seizures. In \cite{truong2018convolutional}, a CNN comprising of three blocks was considered for seizure prediction. Each block incorporated normalization, convolution, and a max-pooling layer. Spectral information was extracted from raw EEG signals utilizing 30-s windows and was further alimented as an input to the CNN. After applying on the benchmark dataset, sensitivity around 81.2\text{\%}, and a mendacious presage rate equal to 0.16 per hour was achieved with CNN. Khan et al.\cite{khan2017focal} assigned a given EEG segment as a preictal, interictal, or ictal class by utilizing a CNN with six blocks to extract features from raw EEG data. Each EEG channel undergoes wavelet transformation at different scales, and its output was finally given as input to CNN. The authors evaluated the same database that attained 0.142 per hour.  \\

A comparison of the proposed work with other previous models on this problem is shown in Table \ref{table:banchmark}.
The fact that the proposed model achieved virtually 98 percent sensitivity and zero False Positive Rate (FPR) is overwhelming. Performance comparison of this work with other machine learning algorithms based on metrics like sensitivity and the false positive rate is demonstrated in Fig \ref{fig:comparision}. All the benchmarks and comparison shown is tested on the CHB-MIT dataset as it is the only publicly available dataset present on EEG right now.

\section{Conclusion}
The algorithms of epileptic seizure prediction have been evolving and increasing inefficiency since the last few decades. With big data analysis, the identification of hidden features from complex EEG signals is made possible. Identification of features in the data that led to better and higher sensitivity was achieved by proposing a system of CNN + LSTM. This model can thus be tested in real-life applications to predict epileptic seizures before so that a patient's quality of life could be significantly improved. After analyzing the model and results, it can be concluded that the model can also be tested with more data sets as in this paper, only the available open (CHB)-MIT data set was used, and more data sets will provide a more clear analysis of the efficiency of the model. It was also realized that there is some scope for improvement in the FPR for the model. In conclusion, the combination of the CNN + LSTM model showed significant improvement. It can be deduced that epileptic prediction algorithms can increase the quality of life and save many lives.

{\small
\bibliographystyle{ieee_fullname}
\bibliography{refs}
}

\end{document}